\documentstyle[referee,psfig,epsf]{mn}
\newif\ifAMStwofonts

\def\simgt{\stackrel{>}{{}_\sim}}

\ifoldfss
  \ifCUPmtlplainloaded \else
    \NewTextAlphabet{textbfit} {cmbxti10} {}
    \NewTextAlphabet{textbfss} {cmssbx10} {}
    \NewMathAlphabet{mathbfit} {cmbxti10} {} 
    \NewMathAlphabet{mathbfss} {cmssbx10} {} 
  \fi
  \ifAMStwofonts
    \ifCUPmtlplainloaded \else
      \NewSymbolFont{upmath} {eurm10}
      \NewSymbolFont{AMSa} {msam10}
      \NewMathSymbol{\upi}     {0}{upmath}{19}
      \NewMathSymbol{\umu}     {0}{upmath}{16}
     \NewMathSymbol{\upartial}{0}{upmath}{40}
      \NewMathSymbol{\leqslant}{3}{AMSa}{36}
      \NewMathSymbol{\geqslant}{3}{AMSa}{3E}

       \let\le=\leqslant
       \let\ge=\geqslant
    \fi
  \fi
\fi 

\ifnfssone
 \newmathalphabet{\mathit}
  \addtoversion{normal}{\mathit}{cmr}{m}{it}
  \addtoversion{bold}{\mathit}{cmr}{bx}{it}
  \newmathalphabet{\mathbfit} 
  \addtoversion{normal}{\mathbfit}{cmr}{bx}{it}
  \addtoversion{bold}{\mathbfit}{cmr}{bx}{it}
  \newmathalphabet{\mathbfss} 
  \addtoversion{normal}{\mathbfss}{cmss}{bx}{n}
  \addtoversion{bold}{\mathbfss}{cmss}{bx}{n}
  \ifAMStwofonts
    \ifCUPmtlplainloaded \else
      \UseAMStwoboldmath
      \makeatletter
      \new@mathgroup\upmath@group
      \define@mathgroup\mv@normal\upmath@group{eur}{m}{n}
      \define@mathgroup\mv@bold\upmath@group{eur}{b}{n}
      \edef\UPM{\hexnumber\upmath@group}
      \new@mathgroup\amsa@group
      \define@mathgroup\mv@normal\amsa@group{msa}{m}{n}
      \define@mathgroup\mv@bold\amsa@group{msa}{m}{n}
      \edef\AMSa{\hexnumber\amsa@group}
      \makeatother
      \mathchardef\upi="0\UPM19
      \mathchardef\umu="0\UPM16
      \mathchardef\upartial="0\UPM40
      \mathchardef\leqslant="3\AMSa36
      \mathchardef\geqslant="3\AMSa3E

       \let\le=\leqslant
       \let\ge=\geqslant
    \fi
  \fi
\fi 

\ifnfsstwo
  \DeclareMathAlphabet{\mathbfit}{OT1}{cmr}{bx}{it}
  \SetMathAlphabet\mathbfit{bold}{OT1}{cmr}{bx}{it}
  \DeclareMathAlphabet{\mathbfss}{OT1}{cmss}{bx}{n}
  \SetMathAlphabet\mathbfss{bold}{OT1}{cmss}{bx}{n}
  \ifAMStwofonts
    \ifCUPmtlplainloaded \else
      \DeclareSymbolFont{UPM}{U}{eur}{m}{n}
      \SetSymbolFont{UPM}{bold}{U}{eur}{b}{n}
      \DeclareSymbolFont{AMSa}{U}{msa}{m}{n}
      \DeclareMathSymbol{\upi}{0}{UPM}{"19}
      \DeclareMathSymbol{\umu}{0}{UPM}{"16}
      \DeclareMathSymbol{\upartial}{0}{UPM}{"40}
      \DeclareMathSymbol{\leqslant}{3}{AMSa}{"36}
      \DeclareMathSymbol{\geqslant}{3}{AMSa}{"3E}

       \let\le=\leqslant
       \let\ge=\geqslant
    \fi
  \fi
\fi 

\ifCUPmtlplainloaded \else
  \ifAMStwofonts \else 
    \def\upi{\pi}
    \def\umu{\mu}
    \def\upartial{\partial}
  \fi
\fi

\title[The Reanalysis of the $ROSAT$ Data of GQ Mus 1983]
{The Reanalysis of the $ROSAT$ Data of GQ Mus (1983) Using White Dwarf Atmosphere Emission
Models}
\author[\c{S}. Balman and J. Krautter]
       {\c{S}. Balman$^1$\thanks{solen@astroa.physics.metu.edu.tr} and
        J. Krautter$^2$ \\
        $^1$Department of Physics, Middle East Technical University, In\"on\"u Bulvar{\i}, Ankara, TR\\
        $^2$Landessternwarte, Koenigstuhl, D-69117 Heidelberg, Germany}

\date{Accepted 
      Received ;
       }
\pagerange{\pageref{firstpage}--\pageref{lastpage}}
\pubyear{2000}

\begin{document}

\maketitle

\label{firstpage}

\begin{abstract}

The analyses of X-ray emission from classical novae during the outburst
stage have shown that
the soft X-ray emission below 1 keV, which is thought to originate from the
photosphere of the white dwarf, is inconsistent with the
simple blackbody model of emission. Thus, $ROSAT$ Position Sensitive
Proportional Counter ($PSPC$) archival data 
of the classical novae GQ Mus 1983 (GQ Mus) have been reanalyzed
in order to understand the spectral development
in the X-ray wavelengths during the outburst stage.
The X-ray spectra are fitted with  the hot white dwarf
atmosphere emission models developed for the remnants of classical novae near
the Eddington luminosity. The post-outburst X-ray spectra of the remnant
white dwarf is examined in the
context of evolution on the Hertzsprung-Russell diagram using
C-O enhanced atmosphere models. 
The data obtained in 1991 August (during the $ROSAT$ All Sky Survey) 
indicate that the
effective temperature is 
kT$_e$$<$54 eV ($<$6.2$\times10^5$ K). The 1992 February data
show that the white dwarf
had reached an effective temperature in the range 38.3-43.3 eV
(4.4-5.1$\times10^5$ K) with an unabsorbed X-ray flux (i.e., $\sim$ bolometric flux) between
2.5$\times 10^{-9}$ and 2.3 $\times 10^{-10}$ erg s$^{-1}$ cm$^{-2}$.
We show that the H burning
at the surface of the WD had most likely ceased at the time of the X-ray observations.
Only the 1991 August data show evidence for ongoing H burning.

\end{abstract}


\section{Introduction}

A classical nova outburst arises from the explosive ignition of
accreted matter
(i.e., thermo-nuclear runaway, TNR) in a cataclysmic binary system
in which a Roche-lobe filling secondary is transferring hydrogen rich
material via an
accretion disk onto the white dwarf (WD) primary.
During the outburst, the envelope of the WD expands to $\sim$ 100
R$_{\odot}$ and 10$^{-7}$ to 10$^{-3}$\ M$_{\odot}$ of material
that reaches the escape velocity is expelled
from the system (Starrfield 1989; Shara 1989; Livio 1994).
The outburst stage usually lasts from a few
months to several years and finally, the system returns to a quiescent
state after outburst.

The TNR models with steady burning of H in an
envelope on the WD surface have been successful in reproducing
the typical nova characteristics in outburst
(Starrfield Sparks, $\&$ Truran 1974; Starrfield et al. 1978; MacDonald 1983;
MacDonald, Fujimoto $\&$ Truran 1985; Starrfield, Sparks, $\&$ Truran 1985;
Prialnik 1986; Truran 1990; Starrfield et al. 1998 and references therein).
According to these models, a gradual hardening of the
stellar remnant spectrum with time past visual maximum is expected
at constant bolometric luminosity and
decreasing radius, as the envelope mass is consumed
(i.e., via H burning and winds).
As the stellar photosphere contracts, the
effective (photospheric) temperature increases
(up to values in the range 1--10 $\times 10^5$ K) and the peak of
the stellar spectrum is shifted from visual to ultraviolet and finally
to the 0.1-1.0 keV X-ray energy band
(\"Ogelman, Krautter, $\&$\ Beuermann 1987; \"Ogelman et al. 1993; MacDonald
1996; Balman, Krautter $\&$ \"Ogelman 1998 and references therein).
The H burning continues at a
constant rate and ceases when a critical mass in the envelope is
reached below which stable hydrogen burning can no longer be established.
The WD photosphere is expected to cool
at constant radius. In addition to the photospheric component, the shocked
nova ejecta
can produce hard X-ray emission above 1.0  keV with
X-ray temperatures $\ge$\ .1 keV ($>$\ 10$^6$ K)
(O'Brien et al. 1994; Orio et al. 1996; Balman et al. 1998).

GQ Mus 1983 (GQ Mus) was discovered in outburst on 1983 January 18.14 UT
(Liller $\&$ Overbeek 1983). It is also the first classical nova
detected in the outburst stage by
$EXOSAT$ (\"Ogelman et al. 1984, 1987). The $EXOSAT$ data was consistent
either with an effective temperature
$\sim$ 25-30 eV (3.0-3.5$\times 10^5$ K)
and a luminosity of 10$^{37}$-10$^{38}$ erg s$^{-1}$ using a
blackbody emission model or a shocked circumstellar material emitting
$\sim$ 1 keV (10$^7$ K) with a luminosity of 10$^{35}$ erg s$^{-1}$
using a thermal plasma emission model.
GQ Mus was observed with the $ROSAT$ $PSPC$ on 1992 February 25-26 and was found to
have an effective temperature of 30 eV
(3.5$\times 10^5$ K)
at near Eddington luminosity using a blackbody model of emission
(L$_E$$\sim$10$^{38}$ erg s$^{-1}$ for 1 M$_{\odot}$ WD star) (\"Ogelman
et al. 1993). This detection $\sim$ 9 years after the outburst
was thought to be a confirmation of the anticipated constant
bolometric luminosity
phase of classical nova outbursts during which stable H burning occurs.
The source was observed by $ROSAT$ on three other occasions following
1992 February.
At the time of the 1993 February 7-11 observation
the temperature of the source was
between 15 and 29 eV (1.7-3.4$\times 10^5$ K) and only an upper limit
of 21 eV (2.4$\times 10^5$ K)
was derived using the 1993 September 3 data (Shanley et al. 1995).
The source flux was unconstrained for all the observations of
GQ Mus which may be attributed to the inappropriate model of emission used to
analyze the X-ray data (i.e., blackbody).
Such problems existed in the first analysis of the X-ray data of
V1974 Cyg 1992 (V1974 Cyg) (Krautter et al. 1996) and the data was reanalyzed
using metal enhanced LTE atmosphere emission models (Balman et al. 1998).

This paper will summarize a spectral reanalysis of the $ROSAT$ $PSPC$ 
archival data of the pointed
and the RASS observations of GQ Mus, 
using better WD atmosphere emission models
developed for H burning hot WDs in classical novae (MacDonald $\&$ Vennes 1991).
We will discuss
the implication of the derived spectral parameters on the standard novae theory
and the characteristics of the WD. 

\section{The Observations And Data}

GQ Mus was observed on five different occasions with the $ROSAT$ $PSPC$ 
four of which were pointed observations;
1992 February 25-26, 1993 February 7-11, 1993 September 3, and 1994 July 6.
The source was also detected during the RASS on 1991 August 5-6 with a
count rate of 0.143$\pm$0.035 c s$^{-1}$ and a low S/N (Voges et al. 1999).
The two other $ROSAT$ $PSPC$ archival data that are presented
were obtained on 1992 March 24-25 (for V838 Her 1991) ,
1993 May 1-2 (for V351 Pup 1991). 
The operational band of $ROSAT$ $PSPC$ is 0.1-2.4 keV and the energy resolution is
($\Delta$E/E)\ $\sim$\ 0.43
at 0.93 keV (Pfeffermann $\&$ Briel 1986).   
The angular resolution of $PSPC$ is 25$^{\prime\prime}$ for on-axis observations.
In the analysis, the data were corrected for vignetting and dead time.
The source and background count rates
were derived using extraction radii in a range
105$^{\prime\prime}$--150$^{\prime\prime}$.
The first 10 channels were excluded from the analysis due to low
statistics and poor
calibration. Table 1 displays important characteristics, count rates
(derived in this analysis), and observation durations of the sources
discussed in this paper.
The X-ray data were
analyzed using EXSAS/MIDAS software package (Zimmermann et al. 1994).

\begin{table*}
\label{1}
\caption{The Count Rates For $ROSAT$ $PSPC$ Pointings Of Detected Galactic Classical Nova And
Important Source Characteristics\ \ $^{*}$ }
\centering
\begin{tabular}{lllllll} \hline\hline
\multicolumn{1}{l}{\ Nova $\&$}  &
\multicolumn{1}{l}{\ E(B-V)} &
\multicolumn{1}{l}{\ Distance } &
\multicolumn{1}{l}{\ t$_3$\ \ $^1$ } &
\multicolumn{1}{l}{\ Time After } &
\multicolumn{1}{l}{\ Count Rate } &
\multicolumn{1}{l}{\ Exposure }\\
Outburst  &  & (kpc) & (days) & Outburst & (c s$^{-1}$) & Time (s) \\
year  & &    &           &  (days) &  &   \\
\hline
GQ Mus 83 (RASS)\ \ $^2$& 0.45$\pm$0.15\ \ $^3$& 4.7$\pm$1.5\ \ $^4$ & 40\ \ $^3$ & 3118 (8.5 yrs)& 0.143$\pm$0.035 & 150 \\
GQ Mus 83 (Pointed) &  & & & 3322 (9.1 yrs) & 0.127$\pm$0.006 & 5848 \\
    &  &   &  & 3641 (10 yrs) & 0.007$\pm$0.002 & 4296\\
    &  &   &  & 3871 (10.6 yrs) & $<$0.003 & 10091 \\
    &  &   &  & 4190 (11.5 yrs) & $<$0.0015 & 3600 \\
QU Vul 84 & 0.5-0.6\ \ $^5$ & 3.5$\pm$1.5\ \ $^6$ & 34\ \ $^4$ &
2340 (6.4 yrs) & 0.003$\pm$0.002 & 10882\\
V838 Her 91 & 0.5\ \ $^7$ & 2.8-5.0\ \ $^7$ & 5\ \ $^5$ & 5 & 0.16$\pm$0.01 & 1235\\ 
     &  &  &  & 365 & 0.005$\pm$0.002 & 4425\\
 V351 Pup 91 & 0.3$\pm$0.1\ \ $^8$ & 4.7$\pm$0.6\ \ $^8$ &
40\ \ $^8$ & 480 & 0.22$\pm$0.01 & 9554 \\
\hline
\end{tabular}
\begin{flushleft}
{({*}) V1974 Cyg 92 is excluded. (see Krautter et al. 1996)}\ \ \
{(1) Time it takes for the nova to decline three magnitudes.}\ \ \
{(2) Voges et al. 1999}\ \ \ 
{(3) Krautter et al. 1984}\ \ \
{(4) \"Ogelman et al. 1987}\ \ \
{(5) Andrillat 1985}\ \ \
{(6) Saizar et al. 1992}\ \ \
{(7) O'Brien et al. 1994}\ \ \
{(8) Orio et al. 1996}\ \ \
\end{flushleft}
\end{table*}

\section{Analysis of the GQ Mus Data with the WD Atmosphere Models}

\subsection{The WD atmosphere models}
The multiwavelength analyses of classical novae have
shown that the ejecta are enhanced in CNO nuclei compared with the sun and
some also show significant enhancement of Ne, Mg and
heavier elements indicating existence of a Ne-O WD.
The enhancement is generally believed to
result from mixing between the WD core material and the accreted
envelope prior to  outburst
(Iben $\&$ Tutukov 1995; Glasner, Livne $\&$ Truran 1997; 
Starrfield et al. 1998).
As a result of the change in the opacity within the stellar remnant
envelope due to
mixing, the emergent stellar remnant
spectrum is expected to differ from a simple blackbody model of emission.
A hot WD spectrum would also show deviations from a simple blackbody spectrum 
(Heise, van Teeseling $\&$ Kahabka 1994).
Thus, we developed a routine to analyze the soft
X-ray spectra of classical novae
using a set of model atmosphere continua provided by MacDonald
(1994, private communication).
The models are hydrostatic LTE atmospheres for a 1.2 M$_{\odot}$
WD at 11 different effective temperatures ranging from 1--10$\times 10^5$ K.
They comprise two representative
compositions: a C-O enhanced atmosphere with C/C$_{\odot}$ $\sim$ 15,
O/O$_{\odot}$ $\sim$ 15 and
Ne/Ne$_{\odot}$ $\sim$ 7; an O-Ne enhanced atmosphere with
C/C$_{\odot}$ $\sim$ 1.2, O/O$_{\odot}$ $\sim$ 14 and
Ne/Ne$_{\odot}$ $\sim$ 81 (MacDonald $\&$\ Vennes 1991).
In addition, H, He and N are also included in the models with N
substantially greater than solar. Other details on the models
and comparison with 
blackbody models of emission can be found in MacDonald $\&$ Vennes (1991).
The atmosphere models were previously used in the X-ray spectral
analysis of the $ROSAT$ data of V1974 Cyg and atmosphere 
models convolved with the
$ROSAT$ response matrix can be find in Balman et al. (1998).
The LTE models are expected to be good approximations to the atmospheres of
remnant hot WDs {\it in the X-ray wavelengths}. This follows from Hartmann $\&$
Heise (1997) where they show that for M$_{wd}$$\ge$ 0.6 M$_{\odot}$ LTE would be
established since collisional ionizations dominate (photoionization). The
radiation field is more strongly coupled to the local temperature (at high
densities) and LTE determines the degree of ionizations and the 
atomic population levels.

\subsection{The spectral analysis}

The visual magnitude at maximum and early evolution of the visual light curve of
GQ Mus was consistent with
shell H burning with a metal enrichment about Z $\sim$ 0.23 and helium abundance
Y $\sim$ 0.4 (Morisset $\&$ Pequignot 1996a).
The spectroscopic observations of the source in the nebular phase
showed evidence for photoionization from a hot radiation source with temperature
$\sim$ 4$\times 10^5$ K (35 eV) that increased in time
(Krautter $\&$ Williams 1989).
We reanalyzed the pointed observations of GQ Mus obtained on 
1991 August 5-6 (RASS data), 1992 February 25-26, and
1993 January 7-11. The other two following pointed observations
on Table 1 were used to derive upper limits on the temperature and flux of the
WD because the source was not detected.
For the analyses, C-O enhanced WD atmosphere models were used
since, the nova was known to be enhanced in CNO nuclei ([CNO/Ne-Fe]$\sim$38.0)
and the Ne/Ne$_{\odot}$ was low (Livio $\&$ Truran 1994; 
Starrfield et al. 1998).

For the 1992 February observation, the effective temperature of the
photosphere was found to lie in the range 38.3-43.3 eV (4.4-5.1$\times 10^5$ K) and the unabsorbed 
X-ray flux (i.e., $\sim$ the bolometric flux) was between 2.5$\times 10^{-9}$\
erg\ s$^{-1}$\ cm$^{-2}$ and 2.3$\times 10^{-10}$\  erg\
s$^{-1}$\ cm$^{-2}$ (at 1$\sigma$ confidence level). Figure 1 shows the fitted
spectrum of GQ Mus using the atmosphere models
that yielded a good fit with $\chi^2_\nu$=1.0. 
The maximum X-ray fluxes were
1$\times 10^{-8}$\ erg\ s$^{-1}$\ cm$^{-2}$ and 3$\times 10^{-8}$\ erg\ s$^{-1}$\ cm$^{-2}$
at 2$\sigma$ and 3$\sigma$ confidence levels, respectively. 

\begin{figure*}
\begin{center}
\psfig{file=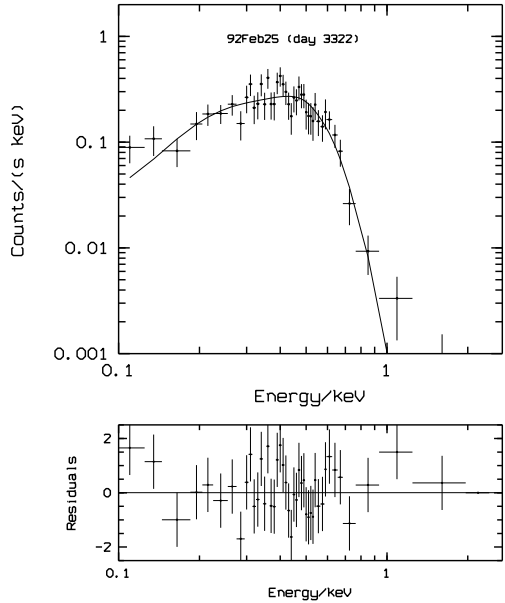,width=16cm,angle=-90}
\caption[]{The X-ray data fitted with the C-O enhanced WD atmosphere
   emission models including neutral hydrogen absorption.
The smooth curve is the PSPC response to the
     model spectrum and the actual PSPC data are indicated with crosses.
   The lower figure shows the residuals between the data and the
     model in standard deviations.}
\end{center}
\end{figure*}

The detected bolometric flux and the
source distance of 4.7$\pm$1.5 kpc yielded a luminosity of about
(0.3-1.2)$\times 10^{37}$\ erg\ s$^{-1}$\ for the central source
(1$\sigma$ range with all the uncertainty in the distance taken into account)
which is a factor of 10 lower than  
the constant bolometric luminosity of GQ Mus found to be $\sim$ 1$\times 10^{38}$\ erg\
s$^{-1}$\ (Hassel et al. 1990) and 1.75$\pm$0.3$\times 10^{38}$\ erg\ s$^{-1}$ (Morisset $\&$ 
Pequignot 1996b). The constant bolometric luminosities detected for GQ Mus
strongly suggest a WD mass M$_{wd}$$\ge$ 0.9 M$_{\odot}$ 
consistent with the large CNO ratio detected for the source as mentioned earlier.
In addition, Starrfield et al. (1996), Prialnik $\&$ Kovetz (1995), 
MacDonald et al. (1985) and Morisset $\&$
Pequignot (1996b) all imply a WD mass 1.1-1.2 M$_{\odot}$ for GQ Mus.
Thus, the choice of the WD mass of 1.2 M$_{\odot}$ in the models used to analyze the ROSAT data
is in good agreement with different estimations.

Tuchman $\&$ Truran (1998) calculate the core mass-luminosity relation for
H burning WDs in classical nova systems as a function of metallicity (i.e., Z). 
This relation yields a constant bolometric luminosity of  9.4 $\times 10^{37}$\ erg\ s$^{-1}$ for a
WD of 0.8 M$_{\odot}$.
We derive a 3$\sigma$ maximum limit of 9$\times 10^{37}$\ erg\ s$^{-1}$
for source distances d$\le$5 kpc using our data (Feb 1992). Higher WD
mass will increase the constant bolometric luminosity at which the H is burned. 
As a result, a scenario where H is still burned can be disregarded at 
3$\sigma$ confidence level for source distances d$\le$5 kpc 
(see also Discussion for further details).

We would like to  add that the fits showed no significant evidence
($\le$2$\sigma$) for a harder X-ray component above 1 keV that existed in
the case of V1974 Cyg (Balman et al. 1998).

The analysis of the 1993 January observation yielded a maximum limit of 35.5 eV
(4.1$\times 10^5$ K)
for the effective temperature of the photosphere and a lower limit of
8.2$\times 10^{-11}$\  erg\
s$^{-1}$\ cm$^{-2}$ for the unabsorbed soft X-ray flux.
The unconstrained flux versus effective
temperature contour derived from the 1993 January data can be explained by
the low count rate which indicates that most of the source spectrum
was outside the $ROSAT$ energy band. Under such conditions, since the
statistical errors are large,
the spectral parameters can not be constrained well.
The upper limits derived from the 1993 September data
on the source temperature and the unabsorbed X-ray flux
are 27 eV (3.1$\times 10^5$ K) and  1.1$\times 10^{-10}$ erg s$^{-1}$\ cm$^{-2}$.
The upper limits are reduced to 
18 eV (2.1$\times 10^5$ K) and 2.2$\times 10^{-11}$ erg s$^{-1}$\ cm$^{-2}$
for the 1994 July data (see also Table 2 in sec. [4]).

Since the 1992 data strongly suggests that the source had already turned off, we also obtained 
the $PSPC$ data taken during RASS in 1991 August to determine better H burning and turnoff time-scales.
The spectral analysis of the data shows that the effective temperature of the photosphere was
kT$<$54 eV ($<$6.2$\times 10^5$ K). 
The spectral parameters of the RASS data were not constrained well due to the low
statistical quality. On the other hand, it shows evidence that the temperature of the
WD was higher at fluxes corresponding to the 1992 data. Since the flux was not constrained, 
it is possible that the WD could be burning H at that time. 

\begin{figure*}
\begin{center}
\psfig{file=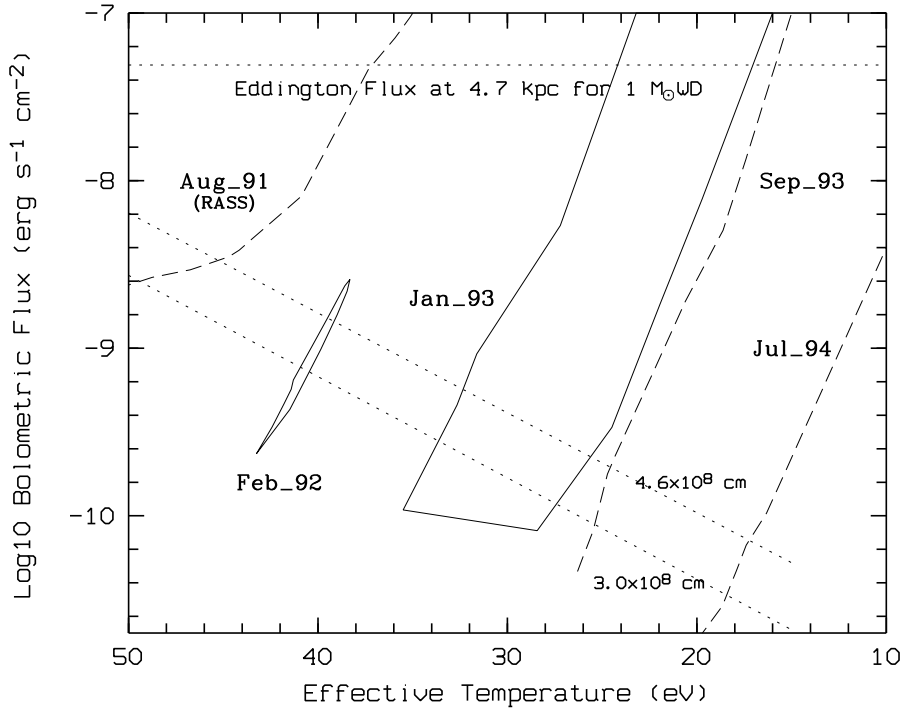,width=16cm,angle=-90}
\caption[]{Effective Temperature versus Bolometric Flux.
The figure shows five 1$\sigma$ confidence
contours derived from the fits with the C-O atmosphere model of emission.
Each contour is labelled by the
observation date. The two dashed lines on the right hand side
show the upper limits for 1993 September
and 1994 July data. The The dashed line on the left hand side show the
maximum flux limit derived from the 1991 August data.
The horizontal dotted line is the Eddington flux
at the distance of the nova (4.7
kpc) for  1M$_{\odot}$ WD. The slanted dotted lines are the constant
radius paths for the cooling stellar
remnant calculated using a 4.7 kpc source distance.}
\end{center}
\end{figure*}

In order to examine the spectral development of the remnant WD
on an effective temperature
versus bolometric flux parameter space (i.e., Hertzsprung-Russell diagram), we prepared confidence
contours from our spectral results. We used the generally accepted range of values for the
column density of neutral hydrogen, N$_H$ $\sim$ (2.0-4.0)$\times 10^{21}$\ cm$^{-2}$ (\"Ogelman
et al. 1987). Figure 2 represents the relative development of the C-O model parameters in time.
The y-axis is the soft X-ray flux from the nova corrected for interstellar N$_H$.
The distance to the
nova does not play any role in the construction of the diagram. The figure shows contours
at 1$\sigma$ confidence level, each labelled by the observation date.
The horizontal dotted line is the Eddington flux at the distance of the nova (4.7 kpc)
for a 1 M$_{\odot}$ WD.
The slanted dotted lines are constant radius paths for the cooling stellar remnant at
3$\times 10^8$\ cm and 4.6$\times 10^8$\ cm. In general, the evolution 
resembles that of a cooling
WD after the H burning ceases.

\section{Discussion and Comparisons}

The X-ray spectral analysis showed that most likely GQ Mus had already 
turned off H burning by the observation in February 1992 
(day 3322 after outburst). We found that at 3$\sigma$ confidence level
the maximum limit of the X-ray luminosity, 9$\times 10^{37}$\  erg s$^{-1}$
at source distances $\le$ 5 kpc, is lower than
the expected luminosity from a WD (M$_{WD}$ $\ge$ 0.8 M $_{\odot}$)
at the constant bolometric luminosity phase of a classical nova evolution.
Since the estimated mass of the WD in GQ Mus is  M$_{wd}$ $\ge$ 0.9
M$_{\odot}$ and the calculated distances are more likely around 2-5 kpcs
(Diaz et al. 1995), the February data is most likely not of 
a H burning WD, but a cooling one.  An earlier 
data (1991 August, day 3118 after outburst) shows evidence for higher WD  
effective temperatures 
($<$54 eV, $<$6.2$\times 10^5$ K) than derived from the 1992 data. 
The unconstrained flux
values of the 1991 August data 
(due to low statistical quality) allows the possibility
for ongoing H burning. Our spectral results are displayed on Table 2.
The analysis of the optical data of GQ Mus with the
photoionization models showed that the source turned off H 
burning before/around 3296 days after 
outburst (26 days before the 1992 X-ray data) which is
in a good agreement with our results (Morisset $\&$ Pequignot 1996b).
The turnoff of H burning is also supported by the IUE data taken in 
1992 April about 48 days 
after the X-ray observation (Starrfield et al. 1996).
Morisset $\&$ Pequignot 1996b
further derived that the source temperature after the turnoff was about 
4.1$\pm$0.6$\times 10^5$ K (for day 3296).
The temperature range, 4.4-5.1$\times 10^5$ K (38.3-43.3 eV), 
derived in this study
for day 3322 after the outburst (1992 February data) is consistent 
with their results.
Overall, there is quite good agreement between
our results on the X-ray, uv and the optical spectral analyses.

GQ Mus was observed with the EXOSAT (i.e., 1984) and the 
ROSAT PSPC (i.e., 1992-1994).
The results are summarized in the introduction. The EXOSAT LE did not have 
energy resolution and the detections were marginal, thus we can not effectively
compare the EXOSAT results with our findings. The ROSAT PSPC results
{\it obtained using simple blackbody model of emission} 
showed a temperature around
30 eV (3.5$\times 10^5$ K) for the 1992 February data and a 
luminosity of 1$\times 10^{38}$ erg s$^{-1}$ consistent with on-going H burning
(\"Ogelman et al. 1993). Shanley et al. (1995) have derived 29-33.5 eV for the
range of temperatures and 1.01-3.77$\times 10^{-8}$ erg s$^{-1}$
cm$^{-2}$ for the X-ray flux using the simple blackbody models. The flux and
temperature are unconstrained at 2$\sigma$ and 3$\sigma$ confidence levels.
The temperature and flux range derived using blackbody model of emission
{\it are not} within our ranges of spectral
parameters at 1$\sigma$-3$\sigma$ confidence level (see also Table [2]). 
We like to note that, there is a slight overlap in flux values only at 
3$\sigma$ confidence level, but the Eddington values are not allowed (see 
sec. [3.2]).  Assuming that the models used 
in this paper are the most up-to-date for remnants of classical novae, they are
more appropriate for analysis 
than simple blackbody model of emission for reasons explained 
in section [3.1]. 
Shanley et al. 1995 derives a range of 15-29 eV for the
WD from the 1993 February data and an upper limit of 21 eV from the 1993
September data {\it using simple blackbody model of emission}. 
Our maximum limits and
upper limits are higher by 30-20 per cent and we derive  an upper limit of 
18 eV for the 1994 July data. In general, the blackbody models underestimate
the effective temperature and over estimate the X-ray flux for GQ Mus.  

In order to investigate more elaborately the turnoff of the nova, we have used the spectral 
parameters derived using the 1991 August and 1992 February data
to construct Figure 3 which shows the contours of acceptable
parameters for the effective temperature in eV
and the photospheric radius in cm. The radii were calculated using the normalization constant 
(R/D)$^2$ and {\it all
the distance uncertainty was taken into account}.
The vertical dashed lines in the figure indicate the radius of a WD with the labelled
mass (of the WD).
The calculations were made using the mass-radius tables of Hamada $\&$ Salpeter (1961) for a C-O WD.
\begin{figure*}
\begin{center}
\vbox{\psfig{file=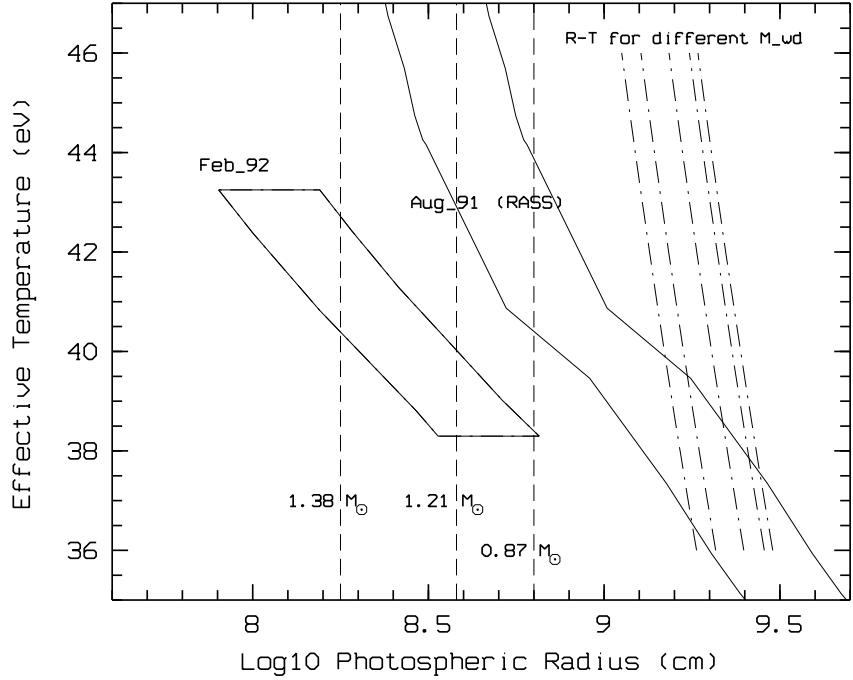,width=16cm,angle=-90}}
\caption[]{Effective Temperature versus Photospheric Radius.
The contours shows the acceptable parameters
for the effective
temperature in eV and the photospheric radius in cm at 1$\sigma$ confidence level.
All the distance uncertainty
is taken into account. The three vertical dashed lines indicate the
mass of the WD that has the particular radii 
(the labels are at the bottom of the lines).
The slanted dot-dashed lines on the right hand side show the R-T relationships for
0.7 M$_{\odot}$, 0.8 M$_{\odot}$, 1.0 M$_{\odot}$, 1.2 M$_{\odot}$ and 1.3M $_{\odot}$
H burning remnant WDs (from left to right).}
\end{center}
\end{figure*}
The dot-dashed slanted lines on the right hand side in Figure 3 show the Radius-Temperature
relationship for 0.7 M$_{\odot}$, 0.8 M$_{\odot}$, 1.0 M$_{\odot}$, 1.2 M$_{\odot}$ and 
1.3 M$_{\odot}$ H burning WDs obtained using the relation of Tuchman $\&$ Truran (1998) 
(L/L$_{\odot}$=52000(M$_{core}$-0.47+0.5Z), where Z=0.23).
{\it The 1.2 M$_{\odot}$ curve (dot-dashed line second from left hand side) 
in Figure 3 overlaps quite well with the calculated R-T relation using our model }
(MacDonald $\&$ Vennes 1991). 
The statistical errors of the 1992 February data are such that there is about 
78 per cent error in the determination of the
photospheric radius in the given range of effective temperatures 38-43 eV. 
However, the R-T relationships for WDs with different masses (as in the
Figure 3) ranging from 0.8M$_{\odot}$-1.3M$_{\odot}$  require statistical errors (of the data in the
photospheric radius) less than
35 per cent so that WD mass can be resolved within $\Delta M$=0.5 M$_{\odot}$. The required 
errors are
less than 10 per cent to resolve the R-T relationships for WD masses with $\Delta 
M$=0.1 M$_{\odot}$ 
(considering the same  effective temperature range of  38-43 eV). As a result, though the
statistics of the data are not bad, the 1992 February
data set and any other ones can not be used to scale the WD mass because it requires quite high
signal to noise.
In addition, since the model fits the 1992 data well with $\chi^2_\nu$=1.0, 
all the model R-T curves plotted in Figure 3 is expected to be within the error contour of 
1992 February data,
displayed in Figure 3 if the WD was still burning the H. {\it For the case of V1974 Cyg most
of the R-T curves above 0.9 M$_{\odot}$ were consistent with the data}.
In the case of 1991 August data, we have found
overlaps between the models  and the maximum limit derived from the data. This is an
evidence that the WD could have been burning the H at that time. However, we caution that
this could also be a manifestation of the low statistical quality of the RASS data. 
The photospheric radii of V1974 Cyg derived using atmosphere emission models are
about a factor of 10 larger than GQ Mus at all times
(V1974 Cyg; R$_{ph}$ $\ge$ 1.9$\times 10^9$ cm; Balman et al. 1998).
The optical spectral analysis using photoionization
models shows that the H burning stops between days 2749 and 3296 after the outburst
(time span is about 1.5 yrs, Morisset $\&$ Pequignot 1996b).
The source is found to decrease in luminosity and increase in temperature for the day 3296.
Thus, the photosphere shrinks about a factor of 4 using the results of
Morisset $\&$ Pequignot (1996b) between these dates. If one assumes an effective
radius of  2.2$\times 10^{9}$ cm for a H burning 1.2 M$_{\odot}$ WD using our model,
the expected radius is then 5.5$\times 10^{8}$ cm which is in good agreement with
our spectral results for the 1992 February data that is about 3322 days after the
outburst. This strongly suggests that the photosphere had almost shrunk to its
original size and the H burning had already ceased sometime ago. The
difference (about a factor of 10) between the photosphere sizes of V1974 Cyg and GQ Mus supports this.
In general, if two hot WDs, one H burning and the other not,
have the same effective temperature, the 
difference between the two will be the {\it g} (surface gravity)
value. The H burning WD will have a lower g (thus, larger photospheric radius) 
compared with a WD
which is hot, but not burning hydrogen.

The cooling time-scale of the WD in X-rays can be calculated assuming an
exponential decay in temperature at constant radius: $T = T_o e^{-t/\tau_c}$
where $\tau_c$ is the cooling time-scale.
Using the spectral results derived in section [3.2], $\tau_c$ is estimated as 3.3 years  by
fitting the declining temperatures (at constant radius) with an exponential decay in time. 
Shanley et al. (1995)
derived about 3-4 yrs for the cooling time-scale using the blackbody emission models
which is similar to our result. This
time scale places the turnoff either in late 1990 or 1991 regardless of
the emission model used assuming that the source is not detected (in X-rays) 
in late 1993 and in the year 1994.
This time-scale is also about 3-4 times longer than the 10 months long  cooling
time-scale detected for V1974 Cyg and it is consistent with the fact that V1974 Cyg is a faster nova than GQ Mus.

The maximum effective temperature
for a post-outburst WD can be given by the empirical relation of MacDonald (1996) as
T$_{max}$ = 6.6$\times 10^{5}$(M$_{wd}$/M$_{\odot}$)$^{1.6}$ K.
The range of effective temperatures derived from the 1992 February data 
yields a lower limit estimate on the WD mass
$\simgt$ 0.8 M $_{\odot}$ since the WD had turned off at that time. 

The H burning time-scale (time passed until turnoff) for GQ Mus is
less than 8.5 years presuming that it just turned off before/around 1991 August.
Over this period, the H-burning rate is predicted to be constant and given by
$\dot M_b = L_P / (X E_H)$.
Assuming that E$_H$ is $\sim$\ 6.4$\times 10^{18}$\ erg\ gr$^{-1}$
(energy obtained by nuclear conversion of hydrogen to helium),
X, the hydrogen mass fraction, is $\sim$ 0.37 (Morisset $\&$ Pequignot
1996a), and L$_P$ is $\sim$ 8.8$\times 10^{38}$\ erg s$^{-1}$ (for 0.8 M$_{\odot}$; Tuchman $\&$ Truran 1998),
then the H-burning rate is $\sim$ 5.8$\times 10^{-7}$\ M$_{\odot}$\ yr$^{-1}$.
The detected H-burning time-scale yields a burned envelope mass of $\sim$\
5.0 $\times 10^{-6}$\  M$_{\odot}$. This value increases to 9.5$\times 10^{-6}$\  M$_{\odot}$
for a 1.2 M$_{\odot}$ WD and thus, corresponds to about $<$ 6-8 per cent of the initial
accreted envelope which is M$_{acc}$ $\sim$ M$_{eject}$ $\sim$
(1.2$\pm$0.2)$\times 10^{-4}$ M$_{\odot}$ (Morisset $\&$ Pequignot 1996a).
Most of the envelope of the WD must have been driven away by an
optically thick wind in the early and/or a radiative wind in the later stages. 
A very similar
result on the burned envelope mass was derived for V1974 Cyg ($<$ 5 per cent; Balman
et al. 1998) and this was attributed to the optically thick wind detected as early as 4-5 days after
outburst. The observed wind phase derived using the optical data of
GQ Mus is $\le$ 0.52 years (Morisset $\&$ Pequignot 1996b).
If this is the case only an optically thick wind with a mass loss rate $\dot M$ $>$ 1$\times 10^{-4}$\
M$_{\odot}$\ yr$^{-1}$  can account for the burned envelope mass and the H burning time-scale.
Such high mass loss rates are consistent with the wind models 
(M$_{wd}\ge$1 M$_{\odot}$ Kato 1997; Kovetz 1998). 

The higher temperatures and smaller radii beyond 1.38 M$_{\odot}$ in Figure 2
translate to an effective emitting region
about 2-6 per cent of the whole WD surface (for 1 M$_{\odot}$ WD)
which can be interpreted as a hot spot
(e.g., as in soft X-ray emitting
regions of Am Her-type systems; G\"ansicke 1998).
The optical and far UV spectra of GQ Mus obtained in years 1989-1994
have also been interpreted as an accreting Am Her-type CV
(e.g.,  double peaked light curve structure in the optical and UV wavelengths;
Diaz $\&$ Steiner 1994; Diaz et al. 1995).
Kahabka (1996) shows the existence of about 30 per cent modulation in the light curve
of the 1992 February data at the orbital
period of GQ Mus which is $\sim$ 85.5 min (Diaz $\&$ Steiner 1989).
This supports that the suggestion that GQ Mus could be accreting in 1992 February however,
the inclination angle of the system is known to be high $\sim$
50$^{\circ}$-70$^{\circ}$ (Diaz $\&$ Steiner 1994)
and thus, occultation by the secondary could easily cause modulations
in the X-ray light curve of the nova. The luminosity derived in section [3.2]
is not a characteristic of AM Her type CVs (eg., 10$^{33-34}$ erg s$^{-1}$ (Cropper 1990))
and such low luminosities could be disregarded at 3$\sigma$ confidence level. The X-ray spectral
evolution of GQ Mus could be simply explained by a cooling WD after a nova 
explosion as expected. An
elaborate scenario that involves accretion can not be directly inferred from 
the X-ray observations
and thus, is beyond the scope of this paper. The 30 per cent modulation
in the light curve can not account for the factor of 10-60 difference
between the observed (unabsorbed) flux and the Eddington Flux.

We also applied the metal enhanced model atmospheres to the X-ray data
of the other Galactic classical nova detected by the $ROSAT$ $PSPC$
in order to compare with the results of the analysis of V1974 Cyg and GQ Mus
(V838 Her 1991, V351 Pup 1991, and QU Vul 1984;
[see Balman 1997 for a review of the analysis]).
Table 2 shows the upper limits on the effective temperatures, photospheric radii,
and bolometric flux of the other two $ROSAT$
detections of Galactic classical novae where we found evidence of soft X-ray emission.
An O-Ne enhanced atmosphere
model of emission was used for the other three novae consistent with the detected
abundance of Ne (Ne/Ne$_{\odot}$ $>$ 40).  
The outburst X-ray spectra of V838 Her, V351 Pup and QU Vul 1984 show
only the harder X-ray component with similar characteristic temperature
and electron density to V1974 Cyg (Lloyd et al. 1991; Orio et al. 1996; Balman $\&$ \"Ogelman 2000).

\begin{table*}
\caption{Spectral Parameters$^1$ for the Soft X-ray (Photospheric)
Components of $ROSAT$ Detections of Classical Novae.}
\begin{tabular}{ccccccc}\hline\hline
\multicolumn{1}{c}{\   Classical Nova }  &
\multicolumn{1}{c}{\   X-ray Obs. Time}  &
\multicolumn{1}{c}{\ kT$_{eff}$ (eV)$^2$ } &
\multicolumn{1}{c}{\ R$_{ph}$ (cm)} &
\multicolumn{1}{c}{\ F (erg s$^{-1}$ cm$^2$)} &
\multicolumn{1}{c}{\ $\chi^2_\nu$ }\\
\hline
GQ Mus 1983  &  8.5 yrs & $<$54 & $>$1.8$\times$10$^8$ & $>$1.3$\times$10$^{-9}$ & 1.0 \\
 &  9 yrs &38.3-43.3 & (1.8-6.3)$\times$10$^8$ & 0.2-2.5$\times$10$^{-9}$ & 1.0 \\
  & 10 yrs & $<$ 35.5 & $>$ 1.8$\times$10$^8$  & $>$ 8.2$\times$10$^{-11}$ & 0.7 \\
  & 10.6 yrs & $<$ 27 & $>$ 1.8$\times$10$^8$ & $<$ 5.3$\times$10$^{-11}$ & 1.4 \\
  & 11.4 yrs & $<$ 18 & $>$ 1.8$\times$10$^8$ & $<$ 2.2$\times$10$^{-11}$ & 0.5 \\
  V838 Her 1991 & 365 days& $<$26 & $<$ 3.8$\times$10$^9$ & $<$ 10$^{-8}$ & 1.2 \\
  V353 Pup 1991 & 480 days& $<$30 & $<$ 3.5$\times$10$^9$ & $<$ 10$^{-8}$ & 1.1 \\
\hline
  \end{tabular}
\begin{flushleft}
{($^1$) ranges and limits are calculated at 1$\sigma$ confidence level}\\
{($^2$) 1eV=1.2$\times$10$^4$K}
\end{flushleft}
\end{table*}
The second observation of V838 Her obtained one year later was a marginal detection (S/N=7).
An effective temperature range of (12.0-26.0) eV (1.4-3.0$\times 10^5$ K for
N$_H$$<$3$\times 10^{21}$ cm$^{-2}$)
was derived at 1$\sigma$ confidence level
which yielded an upper limit of 3.8$\times 10^9$\ cm for the photosphere assuming a WD
emitting at the
Eddington luminosity ($\sim$ 1.7$\times 10^{38}$ erg s$^{-1}$). 
The best-fitting values were an N$_H$ of 0.1$\times 10^{21}$\
cm$^{-2}$ and an effective temperature of 14.0 eV which yielded an effective radius of 3.2$\times
10^8$\ cm for the photosphere at 5 kpc source distance with an
X-ray flux $\sim$ 2.5$\times 10^{-10}$\ erg s$^{-1}$ cm$^{-2}$.
A Raymond-Smith model of
emission (Raymond $\&$ Smith 1977) yielded  $\chi^2_\nu$ values
larger than 3.0 which supported the photospheric emission interpretation.
In order to look for traces of a H burning WD in V351 Pup, we fitted the X-ray spectrum
with a two-component model as in V1974 Cyg (see Balman et al. 1998).
The second component
(Raymond-Smith thermal plasma model) yielded spectral results similar to
the single-component fit.
Using the O-Ne atmosphere emission models, we derived a
range of (25-30) eV (3.0-3.5$\times 10^5$ K) at 1$\sigma$ confidence level
for the temperature of the alleged stellar remnant (1$\times 10^{21}$$\le$N$_H$$\le$
7$\times 10^{21}$ cm$^{-2}$ at 3$\sigma$ confidence level).
This indicated
that the soft X-ray source could be a cooling WD  with an upper limit on the
effective radius $\le$ 3.5$\times 10^9$\ cm using the temperature
estimate above and the
Eddington luminosity ($\sim$ 1.7$\times 10^{38}$ erg\ s$^{-1}$).
Orio et al. (1996) have also 
calculated a similar temperature of (17.0-25.0) eV (2.0-3.0$\times 10^{5}$ K)
for the central source using strong [Ne III] lines in the optical spectra. 

The few examples discussed in this paper indicate that the soft X-ray components
(0.1-10 keV) of classical novae  were not
detected with the $ROSAT$ satellite once the temperature of the stellar remnant was
$<$ 30 eV. It is highly likely that for the case of V838 Her and V353 Pup
the soft component had already subsided at the time of the X-ray observations.
Besides these, Nova LMC 1995
has been detected by $ROSAT$ $PSPC$ as a third soft X-ray source (kT$<$38 eV)
three years after its outburst in 1998 (Orio $\&$ Greiner 1999). 
In addition, the recurrent nova U Sco is detected as a H burning WD (kT$\sim$77.5 eV)
in its recent outburst (Kahabka et al. 1999). Both detections are single
observations in time and an evolutionary scenario can not be drawn.

\section{Conclusions}

The spectral analysis of the soft X-ray data of GQ Mus (0.1-1 keV) show
significant differences from the results obtained using a blackbody
model of emission. The use of metal enhanced atmosphere emission models instead of a
blackbody model of emission to study the soft X-rays from hot WDs is proved once again
to be of great importance.
We find that the source had most likely ceased
H burning prior to February 1992 and was in the cooling stage 
at the time of the X-ray observations indicating that the H burning 
time-scale is $<$ 8.5 yrs. This is also consistent with the
data in the optical and uv wavelengths.   
This study also stresses the distinction between a super soft X-ray source and a H-burning WD as that the 
former does not imply the latter.
The data obtained in 1991 August yields only
a maximum limit on temperature as 54 eV ($\sim$6.2$\times 10^5$ K)
which indicates that the source was more hot in 1991 comprared to 1992, 
supporting the standard nova theory that implies massive WDs in classical nova
systems. We derive
a temperature range of 38.3-43.3 eV (4.4-5.1$\times 10^5$ K) for the central source together with
an unabsorbed X-ray flux ($\sim$ bolometric flux) of (2.5-0.23)$\times 10^{-9}$
erg s$^{-1}$ cm$^{-2}$ for the 1992 February data. All the results are displayed in Table 2.
The lower limit on the mass of the WD is estimated as 
M$_{wd}$ $\simgt$ 0.8M$_{\odot}$. We obtained
3.3 yrs for the cooling time-scale of the classical nova and the source was not detected
in the 1994 July 6 observation as was in the 1993 September 3 observation
(consistent with this time-scale). We calculated that
the burned envelope mass is less than 6-8 per cent of the accreted envelope which suggests that
most of the material was ejected in a wind with $\dot M$ $>$ 1$\times 10^{-4}$\
M$_{\odot}$\ yr$^{-1}$ at the earlier phases of the outburst.
We have not detected a secondary hard component (1-10 keV) expected
as a result of a wind-wind/circumstellar interaction of the ejected material.
However, we suggest that the
cooling time-scale of such a component could be less than 10 yrs since the wind phase was only 0.5 yrs.
GQ Mus is a standing proof (after V1974 Cyg) of the predictions of the standard nova theory where one expects
a hot luminous central star increasing in temperature as the H in the envelope 
is burned and
cooling at constant radius after the turnoff.
There have been few detections of soft X-ray components of classical novae
due to several reasons: 1) Since most novae have been observed in X-rays only once, it is
difficult to predict any evolutionary scenario and detect the existence of 
a H burning hot WD.
2) The length of H burning time scale could be short (i.e., massive WDs) and
the sources could be turned off before the X-ray observation (ie., V838 Her second obs. and V351 Pup).
3) The intrinsic N$_H$ in the shell could be too high
at the time of the X-ray observations and most of the soft X-rays could be absorbed (early
phases of V1974 Cyg). Therefore, monitored target of opportunity observations of classical novae
in X-rays is probably the only way useful information can be obtained 
and used as probes in understanding the nova explosions and standard nova theory
(as it was done for V1974 Cyg [Krautter et al. 1996; Balman et al. 1998]
and Nova Vel 1999 [Orio et al. 1999; Mukai $\&$ Ishida 1999]). 

\section*{Acknowledgments}

We would like to thank J. Tr\"umper, W. Voges and MPE for their kind
readiness to provide us the $ROSAT$ All Sky Survey Data prior to making  them
generally public.  We also thank J. MacDonald
for providing the models that have been used in the analysis 
and for his helpful 
discussions. We thank H. \"Ogelman for his critical reading of the
manuscript. The $ROSAT$ project is supported by the Bundesministerium
f\"ur Bildung, Wissneschaft, Forschung und Technologie (BMBF/DARA) and the 
Max-Planck-Gesellschaft.

\label{lastpage}

\end{document}